\documentclass[12pt]{article}
\usepackage{url,amsmath,amssymb,graphicx}
\begin{document}
\author{J. I. Katz\\
Department of Physics and McDonnell Center for the Space
Sciences\\ Washington University, St. Louis, Mo. 63130}
\title{Dodgeball---Can a Satellite Avoid Being Hit by Debris?}
\maketitle
\begin{abstract}
Can a satellite dodge a collision with untracked orbiting debris?  Can a
satellite dodge collision with a tracked object, making only the avoidance
man{\oe}uvers actually required to avoid collision, despite the
uncertainties of predicted conjunctions?  Satellite-borne radar may
distinguish actual collision threats from the much greater number of near
misses because an object on a collision course has constant bearing, which
may be determined by interferometric detection of the radar return.  A large
constellation of such radars may enable the determination of the ephemerides
of all cm-sized debris in LEO.
\end{abstract}
\section{Introduction}
Large constellations of satellites are planned to be launched in the coming
years, many to provide world-wide internet access, with cumulative totals of
100,000 or more.  As the number of satellites and orbital debris increases,
so does the threat collisions with other orbiting objects.  Collisions not
only damage or destroy the colliding objects, but also create multiple
pieces of debris, increasing future hazards.

How may a satellite may man{\oe}uver to avoid an imminent collision?
Predicted conjunctions, approaches closer than the orbital uncertainties 
risk collision, are comparatively frequent because these uncertainties are
significant.  Avoiding all conjunctions would require a significant
expenditure of man{\oe}uvering fuel, growing as the number of orbiting
objects grows.  Is it possible to man{\oe}uver in the last-minute (or
second) to dodge the rare objects that would actually collide with a
satellite, while ignoring the large majority of those with predicted
conjunctions that are not actually on a collision course?  Is it possible
to dodge collision with objects too small to be tracked from the ground, and
only detected when they approach a satellite?

Approaching objects may be detected by radar.  If detected at sufficient
range and determined to be on a collision course, collision may be dodged by
a displacement transverse to the relative velocity vector.  The tiny
fraction of nearby objects actually on collision courses have constant
bearing, that may be measured interferometrically by comparing the phases
of radar returns received by outriggers a short distance from the protected
satellite.  The combination of radar and aperture synthesis to determine if
a scatterer has constant bearing provides a powerful method of
distinguishing actual collision threats from harmless close approaches.
\section{Detection Ranges}
There are two distinct classes of threats: 
\begin{enumerate}
\item Debris too small to be tracked individually;
\item Bodies that are tracked individually, but whose predicted conjunctions
are uncertain enough that nearly all would be near-misses, whose avoidance
would burden a satellite's man{\oe}uvering resources.
\end{enumerate}
\subsection{Untracked Debris}
\label{untracked}
Untracked debris can arrive from almost any direction, as viewed from the
satellite we wish to protect.  For prograde orbits there will be little from
the direction of motion, but, because of the broad distribution of orbital
inclinations of threat objects, that information does not greatly limit the
approach direction of threats.

Consider a satellite equipped with a radar that radiates isotropically.
This cannot be achieved with a single dipole, whose antenna pattern has
nulls, but can be reasonably approximated with two or three mutually
orthogonal dipoles.  The approximation of isotropic radiation provides an
estimate of what is possible.  If the radar emits an energy $E$, in one or
many pulses or in a modulated continuous transmission, a dipole receiver
with effective area $\lambda^2/2\pi$ will receive, from an isotropic
scatterer with total radar-scattering cross-section $\sigma$ at range $R$,
the returned energy
\begin{equation}
\label{Eret}
E_{ret} = {\sigma \over (4 \pi R^2)^2}{\lambda^2 \over 2\pi} E \approx
9 \times 10^{-19}{\sigma \over 10^2\,\text{cm}^2}\left({\text{1 km} \over R}
	\right)^4 \left({\lambda \over \text{30 cm}}\right)^2 E,
\end{equation}
A high gain antenna would, of course, receive more energy, but this is
precluded by the requirement of isotropic sensitivity---the arrival direction
of an untracked threat is unknown.  We have scaled $\lambda$ to 30 cm
(1 GHz) because objects smaller than 10 cm would have $\sigma \ll
10^2\,$cm$^2$ for waves of this wavelength, while larger objects are likely
to be tracked.  Impact with a 10 cm object, such as would have a radar
cross-section ${\cal O}(10^2)\,$cm$^2$, is likely to be catastrophic and
must be avoided.

For a receiver with noise temperature $T$ and a signal with matched pulse
width $\tau$ and bandwidth $\Delta \nu = 1/\tau$, the signal to noise ratio
is
\begin{equation}
\label{SN}
{S \over N} = {E_{ret} \over k_B T} \approx 28 {E \over \text{1 J}}
\left({\sigma \over 10^2\,\text{cm}^2}\right) {\text{30 K} \over T}
\left({\text{3 km} \over R}\right)^4 \left({\lambda \over \text{30 cm}}
\right)^2,
\end{equation}
where $k_B$ is Boltzmann's constant.  Eq.~\ref{SN} may be inverted to give
the detectability range
\begin{equation}
	\label{Runtracked}
	\begin{split}
	R_{detect} &= \left({E \sigma \over k_B T (S/N)}\right)^{1/4}
	{\lambda^{1/2} \over 2^{5/4} \pi^{3/4}}\\ &= 3.9\left({E \over
	\text{1 J}} {\sigma \over 10^2\,\text{cm}^2}{10 \over S/N}
	{\text{30 K} \over T}\right)^{1/4}\left({\lambda \over \text{30 cm}}
	\right)^{1/2} \text{km}.
	\end{split}
\end{equation}

$R_{detect}$ is determined by the assumed pulse energy $E$, the receiver
noise temperature $T$, the cross-section $\sigma$ and the radar wavelength
$\lambda$.  $T$ is set by the state of the electronic art, but the other
parameters are design choices: $\lambda$ and $\sigma$ set by the design goal
($\lambda \lessapprox \sqrt{2\pi\sigma}$, where the scatterer is described
by an equivalent sphere of radius $a$, $\sigma \approx 2 \pi a^2$ and
$\lambda \lessapprox 2 \pi a$ is required for efficient scattering;
\cite{BW,vdH}) and $E$ by the power available and the radar pulse repetition
frequency (PRF; itself set by the minimum acceptable $R_{detect}$).

Determination of whether an object is on a collision course requires
detection of two radar returns (Sec.~\ref{Bearing}).  For a nominal relative
velocity $V_0 = 10\,$km/s, appropriate to LEO, and $R_{detect} \approx
3\,$km (Eq.~\ref{Runtracked}) $\text{PRF} = \text{1 km}/V_0 = 10$/s and
the mean radar power $P = E \times \text{PRF} = 10\,$W, probably about the
maximum acceptable for an add-on system to proliferated internet satellites.

A collision warning might be issued at a range of $R_{warn} = 2\,$km,
providing a time $\delta t = R_{warn}/V_0 = 0.2\,$s for evasion.  To
displace the satellite by $\delta x = 1\,$m (a nominal satellite size) in
that time would require an acceleration $2 \delta x/(\delta t)^2 \approx 5
g$, likely infeasible.  Because $R_{detect}$ and $\delta t$ scale $\propto
(E\sigma)^{1/4}$, the acceleration scales $\propto (E\sigma)^{-1/2}$.  This
insensitivity means that it is not possible to dodge untracked objects with
$\sigma = {\cal O}(10^2\,\text{cm}^2)$.  Objects with larger radar
cross-sections will be tracked in the forseeable future \cite{SF,LL}.

The prospect of dodging in GEO is more favorable, although the smaller
density of objects there reduces the risk of collision and may make dodging
unnecessary.  In GEO $V_0 \approx 1\,$km/s because the orbital velocities
are less and the range of inclinations is less.  This multiplies $\delta t$
tenfold and reduces the required acceleration to $\approx 0.05 g$, which is
likely to be feasible, even if produced by a point thruster.
\subsection{Tracked Objects}
\label{tracked}
The problem is easier if the threat objects are large enough that they are
individually tracked and their ephemerides are known.  These are not known
accurately enough to predict conjunctions to within the dimensions of
satellites, so avoidance man{\oe}uvers are made whenever there is any risk
of collision, even though it be very small.  The system proposed here could
be used to render nearly all of these man{\oe}uvers unnecessary, provided a
last-chance man{\oe}uver is possible in the rare cases when it is necessary.
This would reduce the expenditure of propellant on man{\oe}uver by a large
factor, as well as reducing any disruption of the satellite's mission by
man{\oe}uver.

Rather than continually searching all nearby space for possible threat
objects, requiring transmission of $\sim 10$ radar pulses per second, it
would only be necessary to resolve the rare predicted conjunctions.  More
energetic radar pulses would be possible, extending the range of detection,
because threatened conjunctions are rare and infrequent.  The energy $E$ of
a radar pulse would be limited only by capacitive energy storage, not by the
mean radiated power.

Tracked objects approach a conjunction from a known direction.  That permits
use of high gain antenn{\ae} for both transmission and reception, and
detection at much greater ranges.  As a nominal example, take a radar
frequency of 22 GHz ($\lambda = 1.36$ cm); at this frequency a dish 40 cm
($30 \lambda$) in diameter has an area $A = 1.3 \times 10^3\,$cm$^2$ and a
gain $G = 4 \pi A/\lambda^2 \approx 40\,$dB in the center of its $2^\circ$
(0.03 radian, $10^{-3}$ sterad) beam.  It is compact and light enough for
inclusion as an add-on system on a satellite, without requiring great
precision of figure ($\lambda/20 = 0.7\,$mm).  It can also be rapidly
slewed, although with conjunctions predicted hours or days in advance this
is unlikely to be necessary.

22 GHz is in the middle of an atmospheric water vapor absorption band, with
typical (but dependent on location, weather and season) attenuation of 20 dB
from the zenith to the ground.  Attenuation minimizes any interference with
other systems, partly by attenuating the the radar radiation and partly
because atmospheric attenuation means that this band is little used. 
Interference might become an issue if a large number of communications
satellites were equipped with collision avoidance systems, both for their
own immediate protection and to avoid multiplication of secondary debris,
although if radars are only turned on when conjunctions are predicted,
infrequent events, this is not expected to be an issue.

An object tracked from the ground is likely to have a scattering
cross-section $\sigma \gtrapprox 10^2\,$cm$^2$, and a satellite (rather than
a debris object) is likely to have $\sigma \sim 10^4\,$cm$^2$.  Replacing
the dipole effective area $\lambda^2/2\pi$ by a dish of area $A$,
Eq.~\ref{Runtracked} becomes
\begin{equation}
\label{Rtracked}
	\begin{split}
	R_{detect} &= \left({\sigma A G \over 16\pi^2} {E \over (S/N)k_B T}
	\right)^{1/4}\\ &\approx 60 \left({\sigma \over 10^2\,\text{cm}^2}
	{A \over 10^3\,\text{cm}^2} {E \over \text{1 J}} {10 \over S/N}
	{\text{30 K} \over T} {G \over \text{40 dB}}\right)^{1/4}\ \text{km}.
	\end{split}
\end{equation}
If the factors in parentheses are unity, $\delta t \approx 6\,$s and the
acceleration required for a displacement $\delta x = 1\,$m would be $\approx
0.005g$, a quite modest value.  Over such times and distances the
gravitational acceleration must be taken into account, but this is not an
obstacle in principle.

If the antenna is correctly pointed when the target is first within the
range $R_{detect}$ (Eq.~\ref{Rtracked}), then comparison of the amplitudes
detected by multiple antenn\ae\ (as discussed in Sec.~\ref{Bearing}, the
proposed system requires at least three antenn{\ae}) can be used to refine
the pointing and track the target.

A large conjunction uncertainty $C \gg 1\,$km would imply that the 40 dB
gain achievable with a 40 cm dish at 22 GHz would be useless---it would not
be possible to point the beam with sufficient accuracy at the detection
range Eq.~\ref{Rtracked} to illuminate the target.  In this case deliberate
defocusing would be necessary to ensure that the target would be within the
beam.  Then the best possible gain would be $G \approx 4 \pi
(R_{detect}/C)^2$ and Eq.~\ref{Rtracked} would become
\begin{equation}
	\begin{split}
		R_{detect} &= \sqrt{{\sigma A \over 4 \pi C^2}{E \over (S/N)
		k_B T}}\\ &\approx 140\sqrt{{\sigma \over 10^2\,\text{cm}^2}
		{A \over 10^3\,\text{cm}^2}{E \over \text{1 J}}{10 \over
		S/N}{\text{30 K} \over T}}\left({\text{1 km} \over C}\right)
		\ \text{km}.
	\end{split}
\end{equation}
\section{Will it Collide?}
\label{Bearing}
Is an object a collision threat?  Radar provides its range and closing
speed, but more is required; most tracked objects that come within a
conjunction uncertainty $C$ will not collide, nor will most untracked
objects within the range of Eq.~\ref{Runtracked}.  As pilots and sailors
know, an object on a collision course keeps a constant bearing as it
approaches.  A single receiver cannot determine bearing, but a minimal form
of aperture synthesis can.  The relative phases among three non-colinear
receivers determine bearing in the two angles required to describe the
direction to the scatterer.  Absolute bearing is required, so any relative
motion (such as rotation around the satellite, if they are deployed
centrifugally) among the receivers must be compensated; in general, this
motion will be known or determined from the phases of a distant source.

It is not necessary to determine the scatterer's actual bearing, but only
to determine if it is changing.  If changing, it is not a collision threat.
For this purpose, the receiving system must comprise a minimum of three
non-colinear receivers.  Radar returns from the receivers are compared
coherently---this is a minimal synthesized aperture or interferometer, with
two visibility functions.

\begin{figure}
\centering
\includegraphics[width=4in]{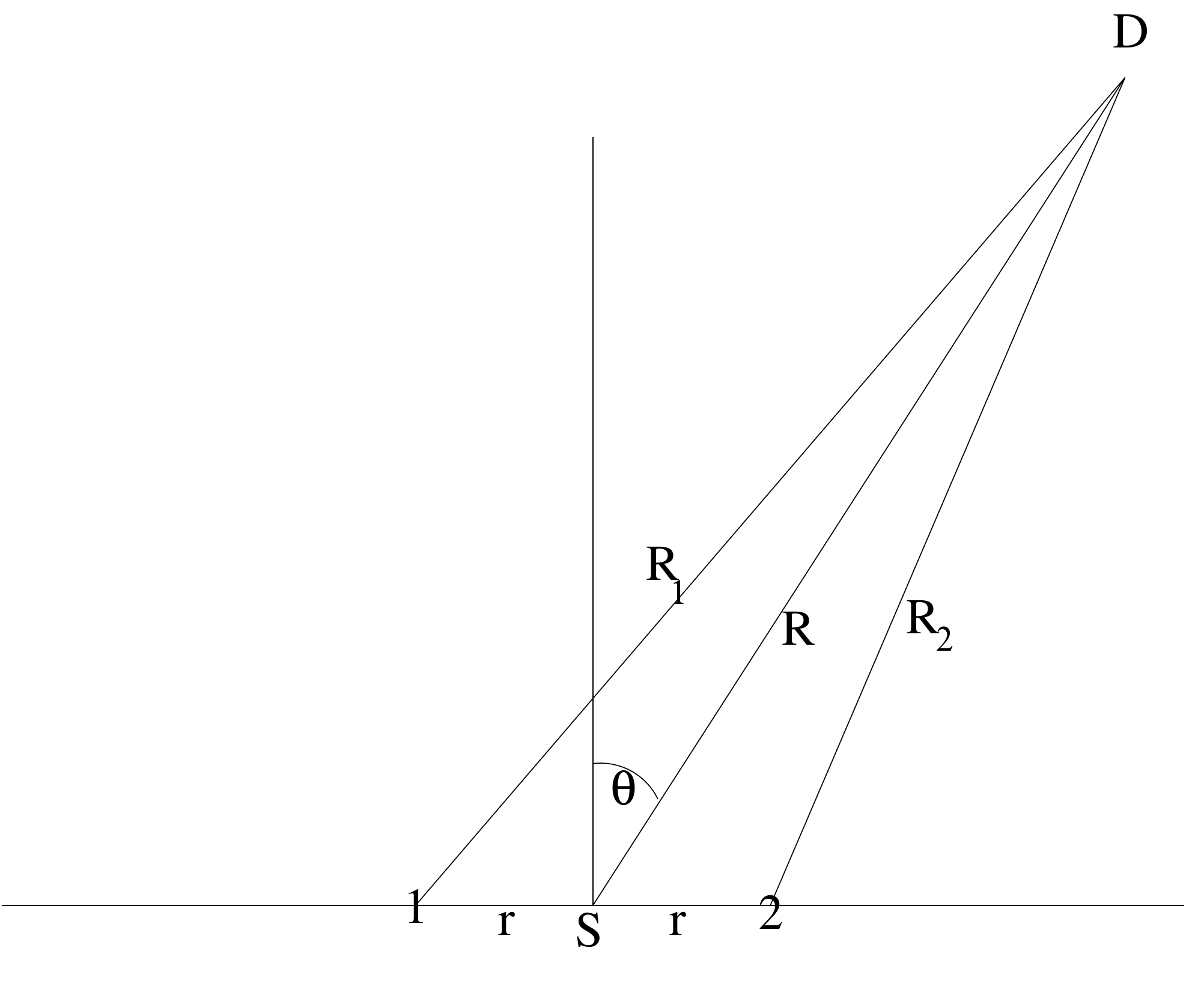}
\caption{\label{dodge1} Path difference between the returns of a radar
signal emitted from a satellite S and received from a threat object D by two
outrigger receivers 1 and 2.  The time derivative of the phase difference
(Eq.~\ref{phidot}) is non-zero unless $\cos{\theta} = 0$, a possibility
dealt with by providing a third receiver not colinear with the first two,
or ${\dot \theta} = 0$, a collision course.}
\end{figure}

Consider (Fig.~\ref{dodge1}) a satellite S threatened by an object, perhaps
debris, D.  The satellite has two colinear outrigger radar receivers at 1
and 2, each a distance $r$ away.  The threat object is at a bearing $\theta$
from the normal to the line connecting the outriggers and satellite.  Radar
signals emitted from the satellite are scattered by the threat object, with
path lengths $R_1$ and $R_2$ to the receivers.  The phase difference between
signals at the two receivers
\begin{equation}
\Delta \phi_{12} = {2 \pi \over \lambda}\left(R_1 - R_2\right) \approx {4 \pi
\over \lambda} r \sin{\theta},
\end{equation}
where $\lambda$ is the radar wavelength and $\theta$ the angle between the
line from satellite to the threat and the line connecting the outriggers
and the satellite.  This result is accurate to first order in $r/R$.

If the threat object is not on a collision course then $\theta$ is changing,
as is the phase difference:
\begin{equation}
\label{phidot}
{\dot{\Delta \phi_{12}}} = {4 \pi \over \lambda} r {\dot\theta} \cos{\theta}.
\end{equation}
${\dot{\Delta \phi_{12}}}$ is nonzero if the object is not on a collision
course.  It is zero if either the object is on a collision course or if
$\cos{\theta} = 0$.  The latter case is dealt with by having a third
receiver, not colinear with the first two.  Then the third receiver and the
satellite will not be colinear with either of the first two receivers
(colinear geometry was illustrated only to ease visualization).  The three
receivers might all be coplanar with the satellite, perhaps on centrifugally
extended booms or wires, equidistant and equally spaced in angle; this is
not necessary, but may be optimal.

\begin{figure}
\centering
\includegraphics[width=4in]{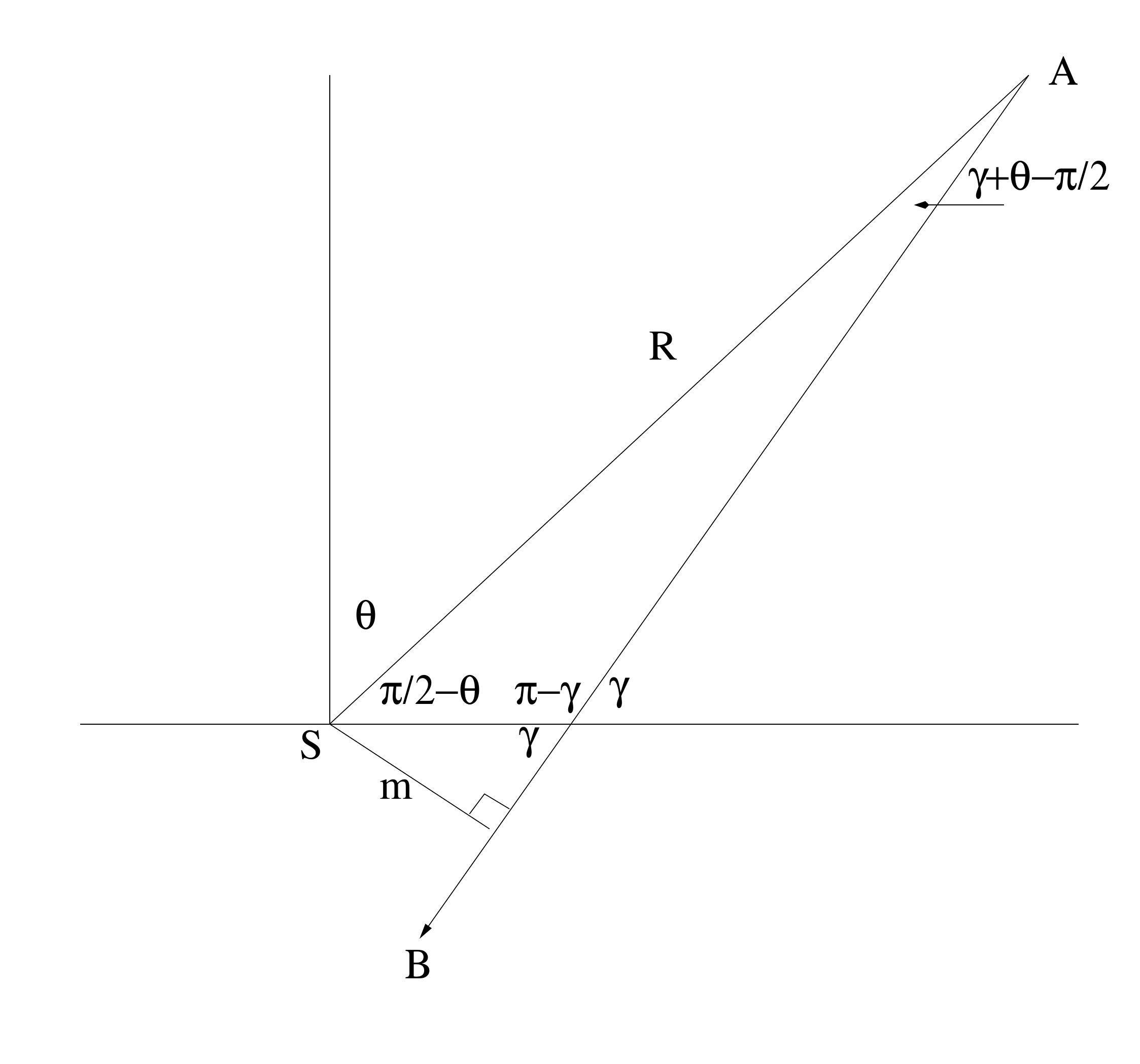}
\caption{\label{dodge2} Encounter geometry, with threat object at range $R$
and miss distance $m$, simplified to the coplanar case.  The horizontal line
joins the receivers and the satellite, as in Fig.~\ref{dodge1}, and the
threat object's trajectory is AB.}
\end{figure}

Fig.~\ref{dodge2} (for simplicity, coplanar geometry has been chosen for
this illustration) shows the threat object's trajectory if the miss distance
is $m$.  This is related to the range $R$ and the angles
\begin{equation}
m = R \sin{\left(\gamma + \theta - {\pi \over 2}\right)},
\end{equation}
where $\theta$ has the same meaning as in Fig.~\ref{dodge1} and $\gamma$ is
the angle between the threat's path and the line joining the receivers.
Differentiating this equation with respect to time ($m$ is constant) we find
\begin{equation}
\label{thetadot}
{\dot\theta} = {1 \over R}{dR \over dt}\cot{\left(\gamma+\theta\right)},
\end{equation}
where $dR/dt \approx V_0$, the threat's velocity in the frame of the
satellite, to first order in $m/R \ll 1$.  $R$ and $dR/dt$ are directly
obtainable from the radar return.

Combining Eqs.~\ref{phidot} and \ref{thetadot}, the rate of change of the
phase difference between the two receivers
\begin{equation}
{\dot{\Delta \phi_{12}}} = {4 \pi \over \lambda}{r \over R}{dR \over dt}
\cos{\theta} \cot{\left(\gamma + \theta\right)} \approx {r \over R}
{m \over R}{4 \pi \over \lambda} V_0 \cos{\theta},
\end{equation}
because for a distant threat object ($m \ll R$) $\gamma + \theta \approx
\pi/2$ and $\cot{\left(\gamma + \theta \right)} \approx m/R$.  Over a time
interval $\Delta t$ (between the returns of two radar pulses) the phase
difference
\begin{equation}
{\Delta \phi_{12}} \approx {4 \pi \over \lambda}{r \over R}{dR \over dt}
\cos{\theta} \cot{\left(\gamma + \theta\right)} \Delta t \approx {4 \pi
\over \lambda} {r \over R} {m \over R} V_0 \cos{\theta} \Delta t.
\end{equation}

Numerically, for a miss distance $m \le 3\,$m (corresponding to a likely
impact on a nominal 3 m satellite), outriggers at $r = 3\,$m, relative
velocity $V_0 = 10$ km/s, range $R = 3\,$km, $\lambda = 1.36\,$cm (22 GHz),
$\cos{\theta} = {\cal O}(1)$ and $\Delta t = 0.1\,$s 
\begin{equation}
\label{dphi}
\Delta \phi_{12} \sim {\dot {\Delta\phi}}\Delta t \sim 10\ \text{rad}.
\end{equation}
This is readily detectable even with a modest signal to noise ratio.
Threat objects with larger $m$ that would miss, would have larger
$\Delta\phi_{12}$ and could be distinguished from actual collision risks.

Eq.~\ref{dphi} is the phase difference in the returns from an object
that risks striking the satellite.  If $\Delta\phi_{12}$ is larger, then, for
the assumed vulnerable radius of $r_{sat} = 3\,$m, we {\it know\/} that the
satellite will not be struck.  Only ${\cal O}(r_{sat}/R_{detect})^2 \sim
10^{-6}$ of the objects entering a 3 km range of detectability really
threaten collision.  The remaining 99.9999\% have $m > 3\,$m and
$\Delta\phi_{12} > 10$ radians and are excluded as threats by measurement of
their $\Delta\phi_{12}$.  For the very large $R_{detect}$ of
Eq.~\ref{Rtracked}, $R_{detect}$ is replaced by the conjunction uncertainty
$C \gg 1\,$m, and the fraction that would actually collide is ${\cal O}
(\text{3 m}/C)^2 \ll 1$.  The problem of objects with small $\cos{\theta}$
is solved by having three (or more) non-colinear outriggers; $\cos{\theta}$
will not be small for some pairs of outriggers. 
\section{Required $\Delta V$}
To dodge a tracked object on a collision course detected at range
$R_{detect}$ (Eq.~\ref{Rtracked}) requires a velocity increment
\begin{equation}
	\label{Vd}
	\Delta V_{dodge} = {r_{sat} V_0 \over R_{detect}},
\end{equation}
where the necessary displacement is $r_{sat}$, taken as the satellite's
(including Solar panels and other appendages) radius.

The alternative is to displace the satellite by the conjunction uncertainty
$C$ each time an object is predicted to come within that distance.
Transverse $\Delta V$ must produce the required displacement in the time
$P_{orb}/2\pi$ because transverse perturbations about a circular orbit are
simple harmonic motion with the orbital period.  It is more effective to 
apply $\Delta V$ along the satellite orbit because there is no restoring
force in that direction; a parallel velocity increment changes the orbital
period and the displacement accumulates secularly over the warning time
$t_{CA}$.  A velocity increment
\begin{equation}
	\label{VC}
	\Delta V_C = C/t_{CA}
\end{equation}
is sufficient to displace the satellite from the conjunction uncertainty
region; it need only be displaced in one dimension to avoid collision.
Displacement along-track can suffice, even if the orbits remain intersecting,
by changing the time at which the satellite passes through the intersection.

For each predicted conjunction that would actually lead to a collision there
are about $(C/r_{sat})^2$ near misses, and the total of the velocity
increments
\begin{equation}
	\label{Vctot}
	\Delta V_{Ctot} = {C \over t_{CA}}\left({C \over r_{sat}}\right)^2.
\end{equation}

The ratio of these two $\Delta V$ is the comparative cost CC of dodging only
those predicted conjunctions that would actually lead to collision:
\begin{equation}
	\label{CC}
	\text{CC} = {\Delta V_{dodge} \over \Delta V_{Ctot}} =
	\left({r_{sat} \over C}\right)^3 {V_0 t_{CA} \over R_{detect}}
	\approx \begin{cases}
		1 \times 10^{-3} & \text{LEO} \\
		3 \times 10^{-6} & \text{GEO},
	\end{cases}
\end{equation}
where the numerical results assume $r_{sat} = 3\,$m, $R_{detect} = 60\,$km
(Eq.~\ref{Rtracked}), $V_0 = 10\,$km/s in LEO and $V_0 = 1\,$km/s in GEO,
$C = 1\,$km in LEO and $C = 5\,$km in GEO and $t_{CA} = 3\,$days in LEO and
$t_{CA} = 10\,$days in GEO \cite{18SCS}.

If the conjunction uncertainty is anisotropic (larger uncertainties in the
along-orbit direction might be expected because errors in the orbital period
produce errors in true anomaly that accumulate in time, just as does the
displacement resulting from along-orbit $\Delta V$), $C$ is a tensor.  It is
likely to have a much greater principal axis along the threat's orbit.

The actual ratios of $\Delta V$ costs are even smaller than those of
Eq.~\ref{CC}:  When the satellite's and threat's orbits intersect at an
angle $\phi \ll 1$ the length of the conjunction uncertainty volume in the
direction parallel to the orbit in Eq.~\ref{VC} is $C \csc{\phi}$.  If the
orbits have nearly the same speed (as for conjunctions between nearly
circular orbits), the $V_0$ in Eq.~\ref{Vd} is multiplied by $\sin{\phi}$.
For these near-parallel conjunctions $\Delta V_{dodge}/\Delta V_C$ is
multiplied by $\sin^2{\phi}$.  For such near-parallel conjunctions it may be
necessary to use transverse $\Delta V$, in which case $t_{CA}$ is replaced
by $P_{orb}/2\pi$ in Eq.~\ref{VC}.  The effect is to reduce CC further.

Unless $C$ is much reduced, last-minute (or last-second) dodging demands
much less propulsion than avoiding all predicted conjunctions because only a
tiny fraction of predicted conjunctions actually need to be dodged.

The extreme ratios of Eq.~\ref{CC} are only applicable if threat objects are
numerous enough, and the satellite is in orbit long enough, that the
probability of a collision that must be dodged is $\ge {\cal O}(1)$, so that
$\ge {\cal O}(C/r_{sat})^2$ predicted conjunctions that would actually be
near misses must be avoided.  In the present environment that is not so;
rather, the probability that any collisions would occur, in the absence of
avoidance, is very small and the number of predicted conjunctions is $\ll
(C/r_{sat})^2$.  CC remains a figure of merit of a Dodgeball system, but has
a different significance.
\section{Survey of Small Debris}
\label{Survey}
The short wavelength radar system described in Sec.~\ref{tracked} can
provide an {\it in situ\/} measure of the density of debris too small to
track individually from the ground, but large enough to pose a significant
threat to satellites.  22 GHz waves have a scattering cross-section $\sigma
\approx 2 \pi a^2 \gtrapprox \lambda^2/2\pi \approx 0.3\,$cm$^2$ for objects
of equivalent radius $a \gtrapprox \lambda/2\pi \approx 2\,$mm
\cite{BW,vdH}.  In lower LEO these may be detectable by powerful
ground-based radars, but in higher orbits they can only be detected {\it in
situ\/}, either by radar or by observation of impact damage.
\subsection{Survey Rate}
Eq.~\ref{Rtracked} provides an estimate of the maximum detection range.  The
beam has a solid angle $\Omega \approx 4 \pi/G\,$sterad.  The volume
surveyed in one pulse
\begin{equation}
	\label{V}
	\begin{split}
		{\cal V} &\approx {4\pi \over 3G}R_{detect}^3
		\approx {4\pi \over 3}\left({\sigma A \over 16 \pi^2}{E \over
		(S/N) k_B T}\right)^{3/4} G^{-1/4}\\ &\approx 3
		\left({\sigma \over 1\,\text{cm}^2}{A \over
		10^3\,\text{cm}^2}{E \over 1\,\text{J}}{10 \over S/N}
		{\text{30 K} \over T}\right)^{3/4} \left({10^4 \over G}
		\right)^{1/4}\text{km}^3.
	\end{split}
\end{equation}
For the nominal parameters (in parentheses) of Eq.~\ref{V}, a single pulse
would detect scatterers with density $> 1\,$km$^{-3}$.  In survey mode
(slewing the beam between pulses by an angle $\ge \Omega^{1/2}$ to avoid
overlap, although objects with a relative speed of $V_0$ move through the
beam in a few tenths of a second) gives a volume sweep rate
\begin{equation}
	\label{sweep}
	\text{SW} = \text{PRF} \times {\cal V}.
\end{equation}
With $\text{PRF} = 10/$s and the nominal parameters of Eq.~\ref{V},
$\text{SW} \approx 30\,$km$^3$/s.  In a year a single radar would sweep a
volume of $10^9\,$km$^3$ and detect objects with a spatial density $\gtrsim
10^{-9}\,$km$^{-3}$.  The minimum detected cross-section $\sigma =
1\,$cm$^2$ corresponds to a radius $a \approx 0.4\,$cm.  Smaller particles
would be detected only if closer, but their density is expected to be
higher, compensating for the smaller $\cal V$ and SW; their detection is
inefficient if $a \ll \lambda/2\pi$ because then $\sigma \ll \pi a^2$, which
effectively limits the survey to objects with $a \gtrsim \lambda/2\pi$.
\subsection{Particle Properties}
Combining the measured $E_{ret}$ and the range $R$ measured by the delay of
the radar return, Eq.~\ref{Eret} permits inference of the scattering
cross-section $\sigma$, and hence of the size of the scatterer $a \approx
\sqrt{\sigma/2\pi}$.

Correlation of the results of {\it in situ\/} radar observations with impact
damage can provide additional diagnostic information.  Radar cross-sections
are largely determined by the greatest linear dimension of an object, while
damage is determined by its kinetic energy (in the frame of the damaged
satellite), and hence by its mass.
\subsection{Ephemerides of Small Debris}
The range to the scatterer would be determined accurately (to $< 1\,$cm) but
its location in the two cross-beam dimensions only to $R_{detect}/G^{1/2}
\approx 0.6\,$km (for the nominal parameters).  With radars operating from a
large number of satellites it might be possible, using the range
measurements and accurate satellite coordinates determined from GPS to
construct accurate ephemerides of even these cm-size scatterers.  With
radars on $10^5$ satellites in future constellations, the entire $3 \times
10^{11}$ km$^3$ of LEO space (altitudes 400--1400 km) could be swept in
about a day.  If the PRF is increased, with a corresponding decrease in $E$
so that the mean radiated power is held constant, SW increases as
PRF$^{1/4}$.

Eq.~\ref{V} shows that survey with a broad angle emitter (like a dipole)
with $G \approx 1$ would be ten times faster than for the assumed $G =
40\,$dB.  This would require an additional transmitter and receiver,
distinct from the high gain system used to predict collision threats, but
would reduce the survey time to a few hours.

In principle, six range measurements from satellites of known location are
sufficient to determine a scatterer's Newtonian orbit, although this is more
complex for the non-Keplerian orbits of the Earth's distorted gravitational
field.  Radar provides not only a range measurement, but, if the pulse is
many cycles long, also a Doppler measurement of radial velocity with respect
to the satellite.  Unfolding orbits from such data, requiring association
of different detections of the same scatterer separated in time when a very
large number of scatterers are observed, is computationally challenging,
but may be possible in principle.
\section{Conclusion}
It is not possible to dodge small untracked objects with radar
cross-sections $\sigma \lessapprox 10^2\,$cm$^2$ with which conjunctions are
not predictable and that may approach from any direction.  It is possible,
with feasible man{\oe}uvers, to dodge larger tracked objects with $\sigma
\gtrapprox 10^2\,$cm$^2$ that have predictable conjunctions.

Search radars on contemplated constellations of $10^5$ satellites may enable
the determination of the ephemerides of all cm-sized debris in LEO space.
This task would be formidable, including computationally, but is not
precluded.
\section{Acknowledgment}
I thank P. Dimotakis, D. Finkbeiner, J. Goodman, K. Pister, C. Stubbs and J.
Tonry for useful discussions.

\end{document}